\documentclass[aps,prc,a4paper,nofootinbib,showpacs,showkeys,
preprintnumbers,superscriptaddress,twocolumn]{revtex4}

\usepackage{amssymb}
\usepackage{bm}
\usepackage{graphicx}
\usepackage{color}
\usepackage[english]{babel}

\newcommand{\eqref}{\ref}

\begin{document}

\title{The influence of chiral chemical potential, parallel electric and magnetic fields on 
the critical temperature of QCD}

\author{M. Ruggieri}\email{marco.ruggieri@ucas.ac.cn}
\affiliation{College of Physics, University of Chinese Academy of Sciences,
Yuquanlu 19A, Beijing 100049, China.}

\author{Z. Y. Lu}\email{luzhenyan11@mails.ucas.ac.cn}
\affiliation{College of Physics, University of Chinese Academy of Sciences,
Yuquanlu 19A, Beijing 100049, China.}

\author{G.~X.~Peng}\email{gxpeng@ucas.ac.cn}
\affiliation{College of Physics, University of Chinese Academy of Sciences,
Yuquanlu 19A, Beijing 100049, China.}
\affiliation{Theoretical Physics Center for Science Facilities, Institute of High Energy Physics, Beijing 100049, China.}


\begin{abstract}
We study the influence of external electric, $E$, and magnetic, $B$, fields parallel to each other, 
and of a chiral chemical potential, $\mu_5$,
on the chiral phase transition of Quantum Chromodynamics. 
Our theoretical framework is a Nambu-Jona-Lasinio model
with a contact interaction. Within this model we compute the critical temperature of chiral symmetry restoration, $T_c$,
as a function of the chiral chemical potential and field strengths. We find that the fields inhibit and $\mu_5$ 
enhances chiral symmetry breaking, in agreement with previous studies.   

\end{abstract}




\maketitle

\section{Introduction}
The existence of chiral anomaly in Quantum Chromodynamics (QCD) ~\cite{Adler:1969gk,Bell:1969ts} 
as well as of nontrivial gauge field configurations at finite temperature \cite{Moore:2000ara,Moore:2010jd}, 
characterized by $F\cdot\tilde{F}\neq0$
where $F$ corresponds to the field stregth tensor and $\tilde{F}$ to its dual, 
has suggested the possibility of observation of chiral magnetic effect  (CME) in relativistic
heavy ion collisions~\cite{Kharzeev:2007jp,Fukushima:2008xe},
as well as of other related effects \cite{Son:2009tf,Banerjee:2008th,Landsteiner:2011cp,Son:2004tq,
Metlitski:2005pr,Kharzeev:2010gd,Chernodub:2015gxa,Chernodub:2015wxa,Chernodub:2013kya,Braguta:2013loa,
Sadofyev:2010pr,Sadofyev:2010is,Khaidukov:2013sja,Kirilin:2013fqa,Avdoshkin:2014gpa}.
A common feature of the aforementioned effects is the excitation of a chiral density, $n_5\equiv n_R - n_L$,
where $n_{R/L}$ correspond to number densities of right-handed and left-handed particles respectively,
induced by the chiral anomaly. Chiral density can also be produced as a Quantum Electrodynamics effect
by coupling quarks to external electric, $E$, and magnetic, $B$, parallel fields for which indeed 
$F\cdot\tilde{F}=E\cdot B\neq0$.
A chiral chemical potential, $\mu_5$,
conjugated to $n_5$~\cite{Ruggieri:2016asg,Ruggieri:2016lrn,Ruggieri:2016cbq,Frasca:2016rsi,
Gatto:2011wc,Fukushima:2010fe,Chernodub:2011fr,Ruggieri:2011xc,Yu:2015hym,Yu:2014xoa,
Braguta:2015owi,Braguta:2015zta,Braguta:2016aov,Hanada:2011jb,Xu:2015vna,
Wang:2015tia,Ebert:2016ygm,
Afonin:2015cla,Andrianov:2012dj,Andrianov:2013qta,Farias:2016let,Cui:2016zqp} 
has been introduced to describe systems in which
chiral density is at equilibrium. 

Because of anomaly,
chiral density is not a conserved quantity in QCD therefore it seems not possible to introduce a chemical potential
for it; however, in a thermal bath microscopic processes
which flip chirality take place on a time scale $\tau$ that is of the order of $1$ fm/c 
around the critical temperature \cite{Ruggieri:2016asg}
(in the quark-gluon plasma phase it is considerably larger \cite{Manuel:2015zpa}),
and for times $t\gg\tau$ it can be easily shown that $n_5$ equilibrates. For example,
in the background of parallel electric and magnetic fields one has \cite{Ruggieri:2016asg}
\begin{equation}
n_5^{\mathrm{eq}} =\frac{q_f^2}{2\pi^2}|eE||eB|\exp\left(
-\frac{\pi M^2}{|q_f eE|}
\right)\tau.
\label{eq:n5eq}
\end{equation}
where $q_f$ corresponds to the electric charge of the flavor $f$ and $M$ to the constituent quark mass.

The above equation shows that a natural framework to study a medium with a chiral density imbalance
is given by quark matter coupled to external parallel electric and magnetic fields: as a matter of fact,
the fields would inject chiral density because of the anomaly; this chiral density would then equilibrate
towards the value given by  Eq.~(\ref{eq:n5eq}), to which $\mu_5$ would be conjugated.
Within this framework it is interesting to discuss the influence of the external fields, as well as of $\mu_5$,
on the thermodynamical properties of quak matter, in particular about chiral symmetry restoration
at finite temperature on which we focus in this article.
Previous studies about this topic can be found in~\cite{Ruggieri:2016asg,Ruggieri:2016lrn,
Babansky:1997zh,Klevansky:1989vi,Suganuma:1990nn,
Klimenko:1991he,Klimenko:1992ch,Krive:1992xh,Gusynin:1994re,Gusynin:1994xp,Cao:2015cka,
D'Elia:2012zw,Cao:2015dya}.

The goal of the present study is the computation of the critical temperature of QCD 
for quark matter in the simultaneous presence of a chiral chemical potential, and 
of external parallel electric and magnetic fields. The model we use is a Nambu-Jona-Lasinio model
with a contact four-fermion interaction responsible for spontaneous chiral symmetry breaking.
Within the model the chiral condensate is rephrased in terms of the constituent quark mass,
whose value is determined by the solution of a gap equation derived within a one-loop
approximation. A similar study has been performed
previously in \cite{Ruggieri:2016lrn} where an expansion in powers
of $\mu_5^2/T^2$, $E/T^2$ and $B/T^2$ has been used to solve the gap equation.
The main purpose of the present study is to improve the results of \cite{Ruggieri:2016lrn}
going beyond the aforementioned expansion. As we discuss in detail throughout the article,
from the qualitative point of view we find no difference with \cite{Ruggieri:2016lrn}:
the parallel $E$ and $B$ act as inhibitors of chiral symmetry breaking lowering the critical temperature,
while the chiral chemical potential tends to increase the latter. From the quantitative point of view
we instead find that the perturbative solution of \cite{Ruggieri:2016lrn} underestimates
the effect of $\mu_5$ on $T_c$, the numerical error being larger for larger values of $\mu_5$ as expected.

One of the main novelties brought by \cite{Ruggieri:2016lrn} has been the computation of the equilibrium
value of the chiral density induced by $E\parallel B$ because of the quantum anomaly; in fact,
the particular configuration of fields leads to the production of $n_5$ which then equilibrates
thanks to chirality flipping processes happening in the thermal bath, see Eq.~(\ref{eq:n5eq}). 
Formally this has been
achieved by solving simultaneously the gap and the number equations, the latter giving the relation
between the equilibrated value of $n_5$ and $\mu_5$. 
In the calculations of \cite{Ruggieri:2016lrn} the value
of $\mu_5$ is therefore computed self-consistently once the value of $n_5$ is known;
on the other hand, in this article we consider a simpler problem,
namely the solution of the gap equation using $\mu_5$ as an external parameter, leaving the
solution of the problem in \cite{Ruggieri:2016lrn} to a near future study.
Therefore the main improvement we bring by  the present study is to go beyond
the small fields/small chemical potential expansion used in \cite{Ruggieri:2016lrn}.
In addition to this, we also consider as a novelty the introduction of the inverse magnetic catalysis
in the model, that has not been taken into account in \cite{Ruggieri:2016lrn}.

The plan of the article is as follows. In Section II we describe the model we use and highlight the main steps
needed to write down the gap equation. In Section III we show and discuss our results.
Finally in Section IV we draw our conclusions.

\section{Thermodynamic potential and gap equation}
We are interested to study quark matter in a background made of
parallel electric, $E$, and magnetic, $B$, fields, in presence of a chiral chemical potential, $\mu_5$.
We assume the fields are constant in time and homogeneous in space; moreover we assume they develop along the 
$z-$direction. 
Our framework is a Nambu-Jona-Lasinio (NJL) model~\cite{Nambu:1961tp,Nambu:1961fr,
Klevansky:1992qe,Hatsuda:1994pi}; the set up of the gap equation and thermodynamic potential 
with a finite chemical potential has been presented in~\cite{Cao:2015dya}, 
therefore we emphasize here only the technicalities involved with the chiral chemical potential
at the same time skipping minor details.

\begin{widetext}
The Euclidean lagrangian density is given by
\begin{equation}
{\cal L}=\bar\psi
\left(
i D\!\!\!\!/ - m_0 
\right)\psi + G\left[(\bar\psi\psi)^2 + (\bar\psi i \gamma_5 \bm\tau\psi)^2 \right],
\end{equation}
with $\psi$ being a quark field with Dirac, color and flavor indices, $m_0$ is the current quark mass
and $\bm\tau$ denotes a vector of Pauli matrices on flavor space. 
The interaction with the background fields is embedded in the covariant derivative 
$D\!\!\!\!/ = (\partial_\mu -i A_\mu \hat q)\gamma_\mu$, where $\gamma_\mu$ denotes
the set of Euclidean Dirac matrices and $\hat q$ is the quark electric charge matrix in flavor space.
In this work we use the gauge $A_\mu = (i E z,0,-B x,0)$. 
Introducing the auxiliary field $\sigma= -2G\bar\psi\psi$ and within the mean field approximation,
the thermodynamic potential in Euclidean spacetime is given by
\begin{equation}
\Omega = \frac{(M - m_0)^2}{4G} - \frac{1}{\beta V}
\mathrm{Tr}\log\left(i\gamma^\mu D_\mu - M - i\mu_5\gamma^4\gamma^5\right),
\label{eq:Omga1}
\end{equation}
where $\gamma^\mu$ denotes the set of Euclidean Dirac matrices, $\beta=1/T$ with $T$ corresponding
to the thermal bath temperature and $V$ is the quantization volume.
The constituent quark mass is $M= m_0 -2G\langle\bar\psi\psi\rangle$ that differs from $m_0$
because of spontaneous chiral symmetry breaking, the latter being related to a nonvanishing
chiral condensate, $\langle\bar\psi\psi\rangle\neq0$. The chiral condensate has its counterpart in QCD, 
but for simplicity we will refer to $M$  keeping in mind that whenever we discuss about the chiral phase transition
in terms of $M$, the decrease of the latter is related to the decreasing of magnitude of the chiral condensate.

\subsection{Traces over color, flavor and Dirac indices}
It turns out it is simpler to compute the derivative $\partial\Omega/\partial M$,
then obtain $\Omega$ by a straightforward integration.
From the above equation we have
\begin{equation}
\frac{\partial\Omega}{\partial M} =
\frac{M-m_0}{2G} - \frac{1}{\beta V}\mathrm{Tr}{\cal S}(x,x),
\label{eq:DeOm1}
\end{equation}
where
\begin{equation}
{\cal S}(x,x) = \mathbf{1}_c \otimes \left(
\begin{array}{cc}
{\cal S}_u(x,x) & 0 \\
0&{\cal S}_d(x,x)
\end{array}
\right),
\end{equation}
with $\mathbf{1}_c$ denoting the identity in color space;
the trace in Eq.~(\ref{eq:GEaa}) is understood over color, flavor, Dirac indices, and
\begin{equation}
{\cal S}(x,x) =\beta V T\sum_n \int\frac{d^3p}{(2\pi)^3}{\cal S}(p).
\end{equation}
The propagator of quark of flavor $f$ in Euclidean momentum space, in case of a static and homogeneous
electromagnetic background and $\mu_5\neq 0$ is given by
\begin{eqnarray}
{\cal S}_f(p) &=&  -i\int\frac{ds}{s^2}\mathrm{det}\left[i q_f F \mathrm{coth}(q_f E s)\right]^{-1/2}
\frac{-(q_f s)^2 I_2}{\mathrm{Im}~\mathrm{cosh}[iq_f s(I_1 + 2i I_2)^{1/2}]}\nonumber\\
&\times & \left[-\gamma\left(1 + \frac{q_f F}{q_f F \mathrm{coth}(q_f F s)}\right)\tilde{p} + M\right]\nonumber\\
&\times & \mathrm{exp}\left(-i M^2 s -i \tilde{p}
\frac{1}{q_f F \mathrm{coth}(q_f F s)}
\tilde{p}
+\frac{i}{2}q_f \sigma F s\right),
\label{eq:fp1}
\end{eqnarray}
where $\tilde{p}$ is a matrix with Dirac and Lorentz structure, namely
\begin{equation}
\tilde{p}=(\omega_n +i\mu_5\gamma_5,\bm p),\label{eq:ptilde}
\end{equation}
and $\omega_n = \pi T(2n+1)$ is the fermion Matsubara frequency. Equation~(\ref{eq:fp1})
can be obtained by the analogous expression for the fermion propagator at finite $\mu$ given
in~\cite{Cao:2015dya} by the replacement $\mu \rightarrow \mu_5\gamma_5$ after noticing
that $\mu_5$ can be interpreted as the fourth component of an Euclidean external axial gauge field,
in the same way $\mu$ is understood as the fourth component of an external Euclidean vector gauge field.

In this section we focus on the computation of $\mathrm{Tr}{\cal S}(x,x)$ in Eq.~(\ref{eq:DeOm1})
at finite $\mu_5$.
To this end we  recall the result~\cite{Schwinger:1951nm}
\begin{eqnarray}
&&\mathrm{exp}\left(\frac{i}{2}q_f\sigma eF s\right) =
\mathrm{cos}(q_f eB s)\mathrm{cosh}(q_f eE s) + i\gamma_5 \mathrm{sin}(q_f eB s)\mathrm{sinh}(q_f eE s)\nonumber\\
&&~~~~~~~~~~
+\gamma_1\gamma_2 \mathrm{sin}(q_f eB s)\mathrm{cosh}(q_f eE s)
 + i \gamma_4\gamma_3 \mathrm{cos}(q_f eB s)\mathrm{sinh}(q_f eE s),\nonumber\\
 &&\label{eq:long}
\end{eqnarray}
which will be useful when traces over Dirac indices will be taken in combination with chirality projectors.
Moreover we have
\begin{equation}
q_f eF \mathrm{coth}(q_f eFs) =
\left(
\begin{array}{cccc}
q_f eE f_1 &0&0&0\\
0 &q_f eB f_2&0&0\\
0 &0&q_f eB f_2&0\\
0&0&0&q_f eE f_1
\end{array}
\right),\label{eq:matrix1}
\end{equation}
with $f_1 = \mathrm{coth}(q_f e E s)$, $f_2 =\mathrm{cot}(q_f e B s)$.

The trace over color and flavor is trivial since the fermion propagator is diagonal in these two spaces.
The trace over Dirac indices is also straightforward: by introducing the chirality projectors
\begin{equation}
{\cal P}_{\pm} = \frac{1\pm \gamma_5}{2},
\end{equation}
and using ${\cal S} = ({\cal P}_{+} + {\cal P}_{-}){\cal S}$,
$[\gamma_5,\sigma^{\mu\nu}]=0$ we are left with the evaluation of
\begin{equation}
\mathrm{Tr}_D\left[ ({\cal P}_{+} + {\cal P}_{-})
(-\gamma {\cal A} \tilde{p} + M)
\mathrm{exp}\left(-i\tilde{p}{\cal B}\tilde{p}\right)
\mathrm{exp}\left(\frac{i}{2}q_f\sigma eF s\right)\right],
\label{eq:7}
\end{equation}
where ${\cal A}$ and ${\cal B}$ can be read from Eq.~(\ref{eq:fp1}) and whose value is not important
in the evaluation of the trace since they do not carry any Dirac structure.
Using $\gamma_5 = {\cal P}_{+} - {\cal P}_{-}$, ${\cal P}_{\pm}e^{i\gamma_5 F} = e^{\pm i F} {\cal P}_{\pm}$,
as well as the fact that the anticommutator of $\gamma_5$ with $\gamma^\mu$ vanishes we can write
Eq.~(\ref{eq:7}) as
\begin{equation}
\sum_{a=\pm 1}\mathrm{exp}\left(-i\tilde{p}_a{\cal B}\tilde{p}_a\right)
\mathrm{Tr}_D\left[
(-\gamma {\cal A} \tilde{p} + M)
{\cal P}_a
\mathrm{exp}\left(\frac{i}{2}q_f\sigma eF s\right)\right],
\label{eq:7b}
\end{equation}
where we have defined
\begin{equation}
\tilde{p}_a = (\omega_n + i a \mu_5,\bm p).
\end{equation}
Because of Eq.~(\ref{eq:long}) we are now left with the
evaluation of the trace of several $\gamma$ matrices combined with chirality projectors. A direct calculation
shows that the only nonzero traces are given by some of the terms proportional to $M$, namely
\begin{eqnarray}
&&\mathrm{Tr} \left[{\cal P}_{\pm}\right] = 2,\\
&&\mathrm{Tr}\left[ \gamma_5{\cal P}_{\pm}\right] = \pm 2,
\end{eqnarray}
that imply we can write the trace over Dirac indices Eq.~(\ref{eq:7}) in the form
\begin{equation}
2M\sum_{a=\pm 1}\mathrm{exp}\left(-i\tilde{p}_a{\cal B}\tilde{p}_a\right)
\left[1 + i\mathrm{sign}(a)\mathrm{sin}(q_f eB s)\mathrm{sinh}(q_f eE s)
\right].
\label{eq:7c}
\end{equation}
The trace of the fermion propagator can therefore be written as
\begin{eqnarray}
&& \frac{1}{\beta V}\mathrm{Tr}{\cal S}(x,x) =2i N_c M\sum_f \sum_{a=\pm1}
\int_0^\infty ds e^{-i s M^2} \nonumber\\
&&\times T\sum_n\int\frac{d^3 p}{(2\pi)^3}
e^{-i\tilde{p}_a {\cal B} \tilde{p}_a}
\left[1 + i\mathrm{sign}(a)\mathrm{tan}(q_f eB s)\mathrm{tanh}(q_f eE s)
\right].
\label{eq:sign}
\end{eqnarray}

\subsection{Integration over 4-momentum}
In Eq.~(\ref{eq:sign}) an imaginary and momentum-independent term proportional to $\mathrm{sign}(a)$ appears,
arising from the term proportional to $\gamma_5$ in Eq.~(\ref{eq:long}) traced with chirality projectors.
In the case of $\mu_5=0$ the trace vanishes trivially; in the case $\mu_5\neq0$ the trace is nonzero,
but still the final contribution of this term to the trace of the propagator vanishes once
the summation over Matsubara frequencies is performed.
In fact performing momentum integration following the same steps depicted
in \cite{Cao:2015dya} we are left, in particular, with the following summation
\begin{equation}
T\sum_n e^{-\tau(\omega_n + ia\mu_5)^2},~~~a=\pm1;
\end{equation}
we have verified that the above summation leads to a real number for all the values of $T$ and $\mu_5$
used in our study, and it is an even function of $a$, hence the summation
over Matsubara frequencies of the term proportional to $\mathrm{sign}(a)$
in Eq.~(\ref{eq:sign}) vanishes. 

Integration over momenta thus leads to
\begin{eqnarray}
&&\frac{1}{\beta V}\mathrm{Tr}{\cal S}(x,x) =
 M\frac{N_c}{2\pi^{3/2}}\sum_f
\int_0^\infty d\tau e^{-\tau M^2}{\cal F},
\label{eq:sign4}
\end{eqnarray}
where  we have defined the function
\begin{equation}
{\cal F} = \frac{qB}{\tanh(qB\tau)}\frac{qE}{\tan(qE\tau)}
\left|\frac{\tan(qE\tau)}{qE}\right|^{1/2}  {\cal T}
\left(\left|\frac{\tan(qE\tau)}{qE}\right|,T,\mu_5\right)
\label{eq:Feb}
\end{equation}
and
\begin{equation}
{\cal T} (\tau,T,\mu_5) = e^{\tau \mu_5^2}T\sum_n e^{-\tau\omega_n^2}\cos\left(2\omega_n \mu_5 \tau\right).
\label{eq:jkl1}
\end{equation}
We remark that the integral over $\tau$ in Eq.~(\ref{eq:sign4}) is assured to be convergent only if
the condition
\begin{equation}
\tau M^2 + \left|\frac{\tan(q E \tau)}{q E}\right|(\pi^2 T^2 - \mu_5^2)>0
\end{equation}
is satisfied: if this is not the case then the Schwinger representation of the propagator
cannot be adopted. Because the $\tan$ function can be arbitrarily large  the above condition
is certainly satisfied for any value of $E$ only if
$\pi^2 T^2 > \mu_5^2$, which limits the domain of the Schwinger representation 
as it happens in the case of finite baryon chemical potential \cite{Cao:2015dya}.

We close this section by noticing that in the $\mu_5=0$ limit the above equation simplifies to
\begin{eqnarray}
&&\frac{1}{\beta V}\mathrm{Tr}{\cal S}(x,x) =
M\frac{N_c}{4\pi^2}\sum_f
\int_0^\infty d\tau e^{-\tau M^2} \nonumber\\
&&\times\vartheta_3
\left(
\frac{1}{2},\frac{i}{4\pi T^2}\left| \frac{qE}{\tan(qE \tau)}\right|
\right) \frac{qB}{\tanh(qB\tau)}\frac{qE}{\tan(qE\tau)},
\label{eq:sign4s}
\end{eqnarray}
in agreement with \cite{Cao:2015dya}, where $\vartheta_3$ denotes the third Jacobi theta-function,
\begin{equation}
\vartheta_3(z,x) = \sum_{n=-\infty}^\infty \mathrm{exp}
\left(
i\pi x n^2 + 2i\pi z n
\right),\label{eq:theta3DEF}
\end{equation}
and we have made use of the inversion formula
\begin{equation}
\vartheta_3(z,x) = \sqrt{\frac{i}{x}}\mathrm{exp}\left(\frac{\pi z^2}{i x}\right)
\vartheta_3\left(\frac{z}{x},-\frac{1}{x}\right).
\end{equation}

\subsection{Regularization}
The behavior of the integrand in Eq.~(\ref{eq:sign4}) for small values of $\tau$ makes the integral divergent,
therefore a regularization scheme has to be adopted. In this article we follow~\cite{Cao:2015dya},
adding and subtracting
the zero field contribution to Eq.~(\ref{eq:sign4}).
In this way, the field-dependent part results finite and therefore independent on the
regularization scheme used; the divergence is confined to the zero field term, which then can be
regularized using any convenient regularization. We use the simple 3-momentum cutoff in this work as it has been
done in~\cite{Cao:2015dya}; we thus have
\begin{equation}
 \frac{1}{\beta V}\mathrm{Tr}{\cal S}(x,x) =-\frac{\partial\Omega_0}{\partial M} +
 M\frac{N_c}{2\pi^{3/2}}\sum_f
\int_0^\infty d\tau e^{-\tau M^2}\tilde{\cal F},
\label{eq:hjkREG}
\end{equation}
where
\begin{equation}
\tilde{{\cal F}} = {\cal F} -\frac{1}{\tau^{3/2}}{\cal T}(\tau,T,\mu_5),
\label{eq:ftilde}
\end{equation}
and $\Omega_0$ denotes the free gas contribution with $\mu_5\neq0$ and $E=B=0$, namely
\begin{eqnarray}
\Omega_0 &=& - N_c N_f \sum_{a=\pm1}\int\frac{d^3p}{(2\pi)^3}\omega_a
- 2 \frac{N_c N_f}{\beta} \sum_{a=\pm1}\int\frac{d^3p}{(2\pi)^3}
\log\left(
1 + e^{-\beta\omega_a}
\right),\nonumber\\
&&\label{eq:Omega0p}
\end{eqnarray}
with $\omega_a = \sqrt{(p+a\mu_5)^2 + M^2}$. Both momentum integrals in Eq.~(\ref{eq:Omega0p})
are understood cutoff at $p=\Lambda$: in this way the ultraviolet divergence in the $T=0$ contribution is removed,
and
the critical temperature for chiral symmetry restoration increases with $\mu_5$
 in agreement with \cite{Yu:2015hym,Ruggieri:2016cbq,Cui:2016zqp,Frasca:2016rsi} as we show in section III.E.

\subsection{Condensation energy and gap equation}
Condensation energy, defined as the real part of $\Omega(M) - \Omega(m_0) $, 
can be obtained by integrating Eq.~(\ref{eq:DeOm1}) over $M$
taking into account Eq.~(\ref{eq:hjkREG}); the integral over $M$ can be performed exactly leading to
\begin{eqnarray}
\Omega(M) - \Omega(m_0) &=& \frac{(M-m_0)^2}{4G} +
( \Omega_0(M) - \Omega_0(m_0))\nonumber\\
&&+ \frac{N_c}{4\pi^{3/2}}\sum_f\left[
\int_{0}^\infty\frac{d\tau}{\tau} ~(e^{-\tau M^2 } - e^{-\tau m_0^2 })
\tilde{{\cal F}}\right],\label{eq:TP2}
\end{eqnarray}
with $\Omega_0$ defined in Eq.~(\ref{eq:Omega0p}).
Because of the $\tan(q_f eE \tau)$ in the denominator of the integrand of Eq.~(\ref{eq:TP2}),
there are simple poles on the integration path for $\tau = \tau_n\equiv n \pi/q_f eE$ with $n=1,2,\dots$.
These poles are treated by adding a small positive imaginary part 
$\tau_n\rightarrow \tau_n - i 0^+$ and using the
Plemelj-Sokhotski theorem to extract the real part
corresponding  to the principal value of the integral \cite{Schwinger:1951nm}.
Equation~(\ref{eq:TP2})
allows to compute the real part, denoted in the following by $\Re$, 
of $\Omega(M) - \Omega(m_0)$ in Eq.~(\ref{eq:TP2}),
namely the difference of free energy between the phase with chiral condensate and the
phase without condensation:
\begin{eqnarray}
\Re\left[\Omega(M) - \Omega(m_0)\right] &=& \frac{(M-m_0)^2}{4G} +
( \Omega_0(M) - \Omega_0(m_0))\nonumber\\
&&+\frac{N_c}{4\pi^{3/2}}\sum_f\left[
\mathrm{PV}\int_{0}^\infty\frac{d\tau}{\tau} ~(e^{-\tau M^2 } - e^{-\tau m_0^2 })
\tilde{{\cal F}}\right].\label{eq:TP3}
\end{eqnarray}

The gap equation is given by $\partial\Re\Omega/\partial M=0$, that is
\begin{equation}
\frac{M-m_0}{2G} = \Re\frac{1}{\beta V}\mathrm{Tr}{\cal S}(x,x),
\label{eq:GEaa}
\end{equation}
Taking into account Eq.~(\ref{eq:hjkREG}) and the principal value prescription
in Eq.~(\ref{eq:TP3}) the regularized gap equation with finite $\mu_5$ and $E||B$
can be written as
\begin{equation}
\frac{M-m_0}{2G}=-\frac{\partial\Omega_0}{\partial M} +
 M\frac{N_c}{2\pi^{3/2}} \sum_f
\mathrm{PV}\int_{0}^\infty d\tau~e^{-\tau M^2 }
\tilde{{\cal F}},
\label{eq:geF1}
\end{equation}
with $\tilde{{\cal F}}$ defined by Eqs.~(\ref{eq:Feb}) and~(\ref{eq:ftilde}).
We use the gap equation to compute the constituent quark mass self-consistently
once temperature, fields and chemical potential are fixed.

\end{widetext}

\section{Results}
In this section we present and discuss our main results. Firstly we consider the case with direct
magnetic catalysis, in order to make a comparison with a previous work. Then we turn to the case
of the model with inverse magnetic catalysis, focusing on the critical line for chiral symmetry restoration
as a function of $\mu_5$ and the external fields strength. 

\subsection{Results with direct magnetic catalysis}

\begin{figure}[t!]
\begin{center}
\includegraphics[width=8cm]{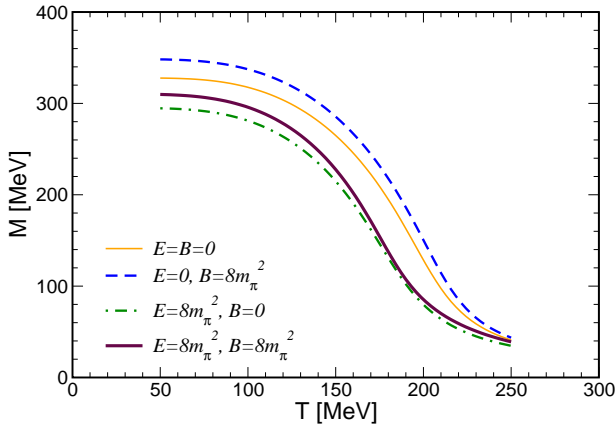}
\end{center}
\caption{\label{Fig:1} Constituent quark mass versus temperature for $E=B=0$ (thin orange solid line),
$E=0$ and $eB=8m_\pi^2$ (blue dashed line), $eE=8m_\pi^2$ and $B=0$ (green dot-dashed line),
$eE=eB=8m_\pi^2$ (thick maroon solid line).}
\end{figure}

In Fig.~\ref{Fig:1} we plot the constituent quark mass versus temperature for
$E=B=0$ (thin orange solid line),
$E=0$ and $eB=8m_\pi^2$ (blue dashed line), $eE=8m_\pi^2$ and $B=0$ (green dot-dashed line),
$eE=eB=8m_\pi^2$ (thick maroon solid line).
These results have been obtained by the solution of the gap equation (\ref{eq:geF1});
they are in qualitative agreement with \cite{Ruggieri:2016lrn}. In fact,
the magnetic field in this model acts as a catalyzer of chiral symmetry breaking,
inducing both a larger value of the constituent quark mass and a larger critical temperature, $T_c$,
the latter identified with the inflection point of the curve $M(T)$.
On the other hand, the pure electric field acts as an inhibitor of chiral symmetry breaking:
quark mass and critical temperature with $E\neq0$ are lower than the ones obtained at zero field.
Finally, the combined effect of $E\parallel B$ is still to inhibit chiral symmetry breaking. In comparison with 
the results at $B=0$ and finite $E$, see solid indigo and green dot-dashed curves in  Fig.~\ref{Fig:1},
quark mass in the case of $E\parallel B$ is a little bit larger because of the magnetic catalysis;
but even if $E=B$ the net effect of the fields is to lower both $M$ and $T_c$ with respect to the zero fields case.

\begin{figure}[t!]
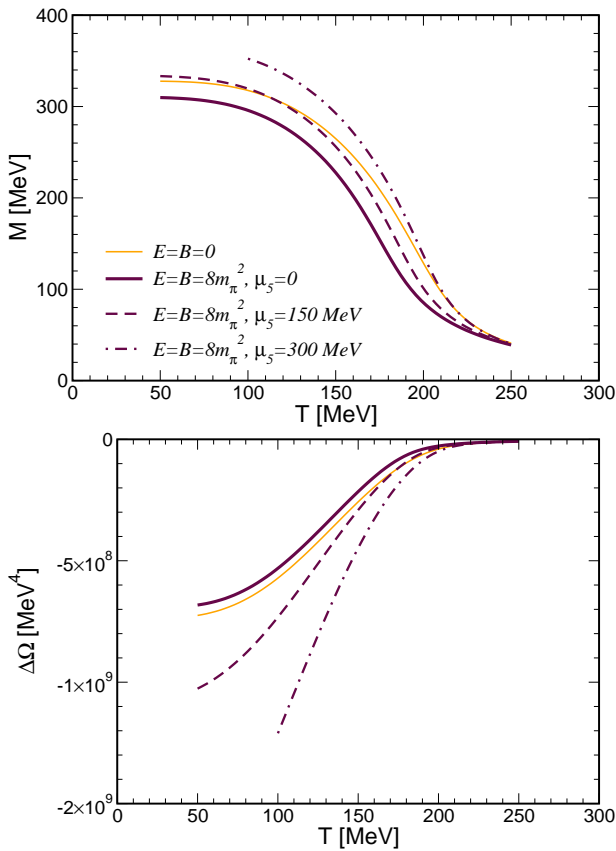

\begin{center}
\includegraphics[width=8cm]{MvT1_5.eps}\\
\includegraphics[width=8cm]{CEvT1_5.eps}
\end{center}
\caption{\label{Fig:2} Upper panel: Constituent quark mass versus temperature for $E=B=0$ (thin orange solid line),
$eE=eB=8m_\pi^2$ and $\mu_5=0$ (thick maroon solid line),
$eE=eB=8m_\pi^2$ and $\mu_5=150$ MeV (maroon dashed line),
$eE=eB=8m_\pi^2$ and $\mu_5=300$ MeV (maroon dot-dashed line).
Lower panel: condensation energy versus temperature (lines convention is the same of the left panel).}
\end{figure}

In the upper panel of Fig.~\ref{Fig:2} we plot the constituent quark mass versus temperature
for $E=B=0$ (thin orange solid line),
$eE=eB=8m_\pi^2$ and $\mu_5=0$ (thick maroon solid line),
$eE=eB=8m_\pi^2$ and $\mu_5=150$ MeV (maroon dashed line),
$eE=eB=8m_\pi^2$ and $\mu_5=300$ MeV (maroon dot-dashed line).
On the right panel of Fig.~\ref{Fig:2} we plot the condensation energy defined in Eq.~(\ref{eq:TP3}).
As expected the effect of $\mu_5$ is to favour chiral symmetry breaking:
at a given temperature both $M$ and the magnitude of the condensation energy
increase  with $\mu_5$.

The results shown on the left panel of Fig.~\ref{Fig:2} allow to discuss the interplay
of $\mu_5$ and the fields on the critical temperature.
In the case $\mu_5=0$ and $E\neq0$, $B\neq 0$
the critical temperature is smaller than the one in the case $E=B=0$:
the combined effect of the parallel electric and magnetic fields is to inhibit chiral symmetry breaking.
On the other hand, increasing the value of $\mu_5$ keeping the values of the fields fixed,
the inflection point of $M$ is pushed towards
higher values of temperature, implying $T_c$ increases with $\mu_5$. It is interesting that
for $\mu_5=150$ MeV the critical temperature is still smaller than the one for $E=B=0$ and $\mu_5=0$,
while if we take $\mu_5=300$ MeV then $T_c$ is larger than the one with zero fields.

\subsection{Comparison with the perturbative calculation}

\begin{figure}[t!]
\begin{center}
\includegraphics[width=8cm]{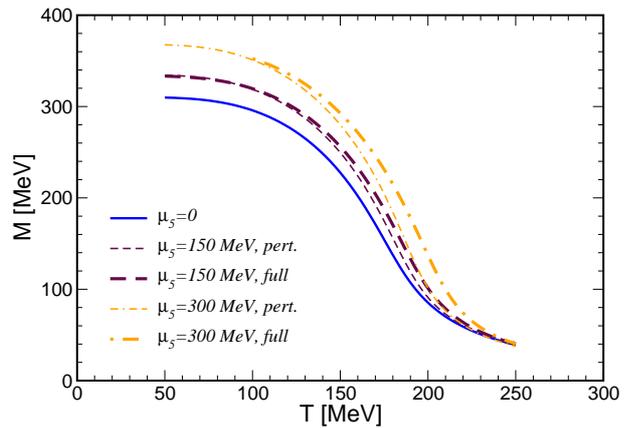}
\end{center}
\caption{\label{Fig:3} Constituent quark mass versus temperature for
$eE=eB=8m_\pi^2$ and several values of $\mu_5$. Blue solid line corresponds to $\mu_5=0$;
maroon dashed lines to $\mu_5 = 150$ MeV, and orange dot-dashed lines to $\mu_5 = 300$ MeV.
Thick lines correspond to the solution of the full gap equation, while thin lines to the perturbative solution
of \cite{Ruggieri:2016lrn}.}
\end{figure}

In this section we compare the results obtained in the present work with those of \cite{Ruggieri:2016lrn} where
a perturbative approach has been used to solve the gap equation in presence of the chiral chemical potential.
In particular, the results of \cite{Ruggieri:2016lrn} have been obtained neglecting any $\mu_5$ dependence
in the field-dependent contribution of the gap equation; since in the present work such a dependence is taken
into account it is possible to compute its effect on the chiral phase transition.

In Fig.~\ref{Fig:3} we plot $M$ versus temperature for
$eE=eB=8m_\pi^2$, that has been also considered in \cite{Ruggieri:2016lrn},
and several values of $\mu_5$. Blue solid line corresponds to $\mu_5=0$;
maroon dashed lines to $\mu_5 = 150$ MeV, and orange dot-dashed lines to $\mu_5 = 300$ MeV.
Thick lines correspond to the solution of the full gap equation, while thin lines to the perturbative solution
of \cite{Ruggieri:2016lrn}. We notice that although the qualitative picture is unchanged in turning from the
perturbative to the full solution, the former underestimates the response of $M$ to $\mu_5$ at temperatures
around the chiral crossover, resulting in the underestimate
of the shift of the critical temperature induced by $\mu_5$. The results in Fig.~\ref{Fig:3} show that
the discrepancy between the perturbative and the full solution is not very important for $\mu_5=150$ MeV,
becoming however sizeable at $\mu_5=300$ MeV.
This shows that quantitatively the results of \cite{Ruggieri:2016lrn} should be taken with a grain of salt
in the case of moderate values of the fields and $\mu_5$.

\subsection{Results with inverse magnetic catalysis}
In this subsection we discuss our results obtained taking into account the inverse magnetic catalysis (IMC)
of chiral symmetry breaking, that we implement in the model calculation by introducing a $B$-dependent
coupling constant \cite{Ferreira:2014kpa} tuned to fit Lattice data about critical temperature \cite{Bali:2011qj},
namely
\begin{equation}
G(\zeta) = G_0 \left(
\frac{1 + a \zeta^2 + b \zeta^3}{1 + c \zeta^2 + d \zeta^4}
\right),\label{eq:IMC1}
\end{equation}
with $G_0$ denoting the coupling at $B=0$ and $\zeta = eB/\Lambda_{\mathrm{QCD}}^2$;
values of the parameters are $a=0.0109$, $b=-1.013\times 10^{-4}$, $c=0.022$, $d=1.846\times 10^{-4}$
and $\Lambda_{\mathrm{QCD}} = 300$ MeV.

\begin{figure}[t!]
\begin{center}
\includegraphics[width=8cm]{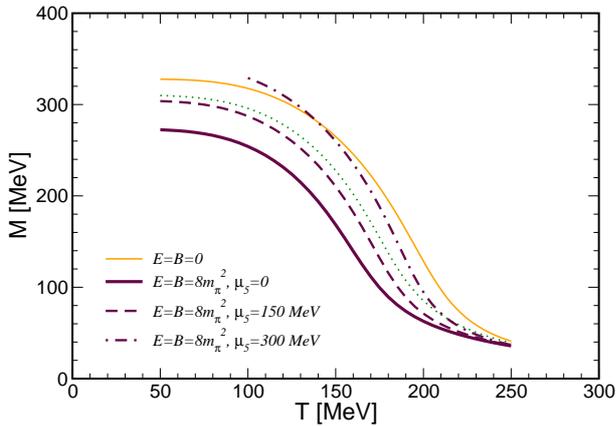}
\end{center}
\caption{\label{T-M-IMC} Constituent quark mass versus temperature for $E=B=0$ (thin orange solid line),
$eE=eB=8m_\pi^2$ and $\mu_5=0$ (thick maroon solid line),
$eE=eB=8m_\pi^2$ and $\mu_5=150$ MeV (maroon dashed line),
$eE=eB=8m_\pi^2$ and $\mu_5=300$ MeV (maroon dot-dashed line). IMC has been 
taken into account. Green thin dotted line corresponds to quark mass for $eE=eB=8m_\pi^2$, $\mu_5=0$
and direct magnetic catalysis already shown in Fig.~\ref{Fig:2}.}
\end{figure}

In Fig.~\ref{T-M-IMC} we plot the constituent quark mass versus temperature with IMC for 
 $eE=eB=8 m_{\pi}^2$. In particular, thick maroon solid line
corresponds to $\mu_5=0$,  maroon dashed line to $\mu_5=150$ MeV,
maroon dot-dashed line to $\mu_5=300$ MeV. For comparison we have shown by the thin solid orange line
the solution of the gap equation at $E=B=0$ and $\mu_5=0$, and by the thin dotted green line
the quark mass for the case $eE=eB=8m_\pi^2$, $\mu_5=0$
and direct magnetic catalysis already shown in Fig.~\ref{Fig:2}.
As expected the behavior of quark mass versus $\mu_5$ is qualitatively not affected by IMC.
The main effect of IMC is to lower considerably the quark mass in comparison with the case
of direct catalysis; the same effect is measurable on the critical
temperature, as it can be noticed comparing the location of the inflection points of the solid maroon 
and dotted green lines in Fig.~\ref{T-M-IMC} (we show results about the critical
temperature in section III.E).

\begin{figure}
  \begin{center}
  \includegraphics[width=9cm]{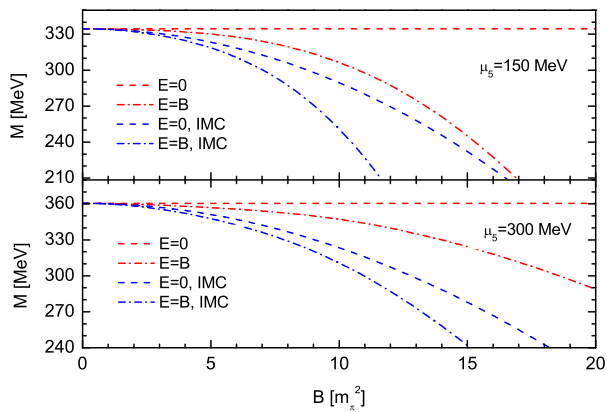}
  \end{center}
  \caption{Constituent quark mass versus magnetic field with (blue lines)
  and without (red lines) the inverse magnetic catalysis for
  $E=0$ (dashed lines) and $E=B$ (dash-dotted lines). Upper panel corresponds to $\mu_5=150$ MeV,
  lower panel to $\mu_5=300$ MeV. Both panels correspond to $T=100$ MeV.}\label{B-M-compri-IMC-No}
\end{figure}

For completeness in Fig.~\ref{B-M-compri-IMC-No} we plot the constituent quark mass 
as a function of magnetic field strength
with (blue lines) and without (red lines)
IMC for $E=0$ (dashed lines) and $E=B$ (dash-dotted lines); upper panel corresponds to $\mu_5=150$
and lower panel to $\mu_5=300$ MeV. All the results are shown for $T=100$ MeV.
As expected, for the cases $E=0$
with magnetic catalysis the magnetic field slightly increases the constituent quark mass.
However taking IMC into account $M$ decreases with $B$.
Then, combining $E$ and $B$ causes the constituent quark mass to drop faster.
Finally, we notice that comparing the results obtained for the two values of $\mu_5$,
increasing $\mu_5$ results in a larger $M$ as it should
since $\mu_5$ enhances chiral symmetry breaking.

\subsection{Relation between $\mu_5$ and $n_5$}
We close this section by commenting briefly on the relation between $n_5$ and $\mu_5$
in presence of constant $E$ and $B$. We limit ourselves only to the case in which inverse magnetic catalysis
is taken into account in the calculation, since it should be the case closest to actual QCD.

\begin{figure}
\begin{center}
  \includegraphics[width=8cm]{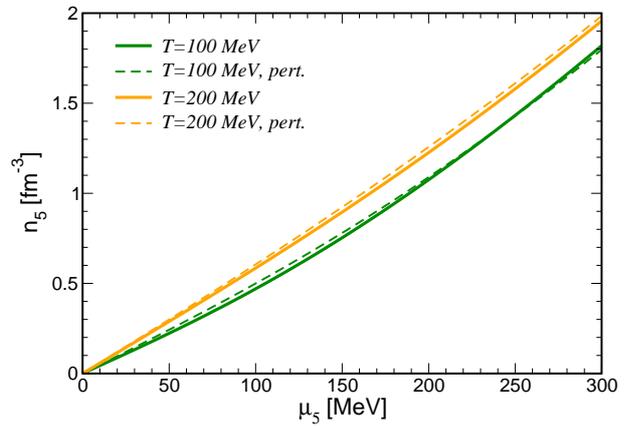}
  \end{center}
\caption{Chiral density versus chiral chemical potential for two temperatures: green lines correspond
to $T=100$ MeV and orange lines to $T=200$ MeV. Solid lines denote $n_5$ obtained
by using the full thermodyamic potential, dashed lines to the perturbative calculation
of \cite{Ruggieri:2016lrn} in which only $\Omega_0$ has been used.
Moreover we have set $eE=eB=8m_\pi^2$.}\label{T-n5-IMC}
\end{figure}

Chiral density is defined as $n_5=-\partial\Omega/\partial\mu_5$ or, by virtue of Eq.~(\ref{eq:TP3}), as
\begin{eqnarray}
n_5 &=& -
\frac{\partial\Omega_0(M)}{\partial\mu_5}\nonumber\\
&& -\frac{N_c}{4\pi^{3/2}}\sum_f\left[
\mathrm{PV}\int_{0}^\infty\frac{d\tau}{\tau} ~e^{-\tau M^2 }
\frac{\partial\tilde{{\cal F}}}{\partial\mu_5}\right].\label{eq:N5_1}
\end{eqnarray}
In right-hand side of above equation the first term is field-independent; we find that numerically it gives the
main contribution to $n_5$. 
The second term corresponds to the
field-dependent contribution that has been ignored in previous calculations \cite{Ruggieri:2016lrn}. 
One of the goals of \cite{Ruggieri:2016lrn} was to compute the 
value of $\mu_5$ at equilibrium once $n_5$ was known. However, in that work the relation
between $n_5$ and $\mu_5$ has been computed perturbatively, 
that is considering only  $\Omega_0$ in Eq.~(\eqref{eq:N5_1})
at the lowest order in $\mu_5^2/T^2$. In this work we want to check quantitatively how good is
the perturbative approximation for $n_5(\mu_5)$ used in \cite{Ruggieri:2016lrn}. 
This is a quick way to estimate the accuracy of the perturbative calculation
of the equilibrium value of $\mu_5$ done in \cite{Ruggieri:2016lrn}.

In Fig.~\ref{T-n5-IMC} we plot $n_5$ versus $\mu_5$ 
for two temperatures: green lines correspond
to $T=100$ MeV (that is below $T_c$) and orange lines to $T=200$ MeV
(that is above $T_c$). Solid lines denote $n_5$ obtained
by using the full thermodyamic potential, dashed lines to the perturbative calculation
of \cite{Ruggieri:2016lrn} in which only $\Omega_0$ has been used.   
Values of the fields are $eE=eB=8m_\pi^2$.
Data have been obtained by fixing temperature and field strengths, then varying $\mu_5$
and computing $n_5$ by virtue of Eq.~(\ref{eq:N5_1}) with $M$ computed self-consistently by solving
the gap equation at finite $\mu_5$.
We find that the fields contribution to $n_5$ is numerically not very important:
for example, for $T=100$ MeV  and an average value of the quark mass $M=330$ MeV
in the range $(0,300)$ MeV of $\mu_5$, we have from Eq.~(\ref{eq:n5eq}) 
$n_5^{\mathrm{eq}}\approx 0.01$ fm$^{-3}$ which would
correspond to $\mu_5^{\mathrm{eq}}\approx 5$ MeV for both the approximate and the full solution;
similarly, for $T=200$ MeV  and an average value of the quark mass $M=100$ MeV
we find  $n_5^{\mathrm{eq}}\approx 0.16$ fm$^{-3}$ which would
correspond to $\mu_5^{\mathrm{eq}}\approx 15$ MeV for both the approximate and the full solution.
Therefore in both cases shown in Fig.~\ref{T-n5-IMC} we find that the relative error induced by replacing
the full thermodynamic potential with the zero field one, concerning the relation $n_5(\mu_5)$,
is negligible.

\subsection{The suggested phase diagram}
\begin{figure}[t!]
\begin{center}
\includegraphics[width=8cm]{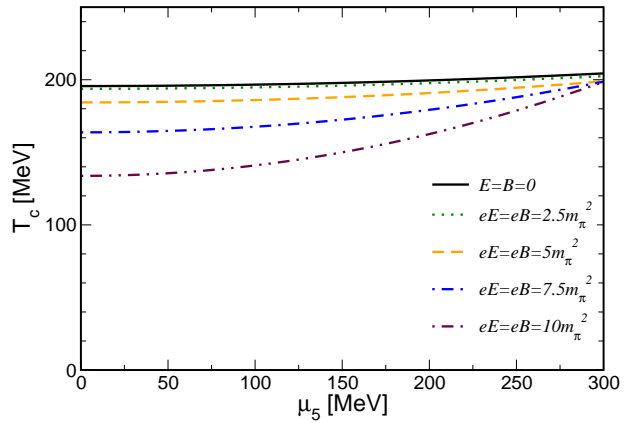}
\end{center}
\caption{\label{Fig:5}  Critical temperature versus $\mu_5$ for several values of $E$ and $B$.
From upper to lower curve we have shown fata for $E=B=0$ (thin black solid line),
$eE=eB=2.5m_\pi^2$ (green dotted line),
$eE=eB=5m_\pi^2$ (orange dashed line),
$eE=eB=7.5m_\pi^2$ (blue dot-dashed line) and $eE=eB=10m_\pi^2$ (maroon dot-dot-dashed line).
Inverse magnetic catalysis has been taken into account in the calculations.
}
\end{figure}

In Fig.~\ref{Fig:5} we plot  the critical temperature versus $\mu_5$
for several values of $E$ and $B$. Inverse magnetic catalysis has been considered in the calculations
by introducing a $B$-dependence of the NJL coupling according to Eq.~(\ref{eq:IMC1})
(results with direct catalysis are qualitatively similar, hence we do not show them).
Since bare quark mass is finite the chiral transition is actually a smooth crossover, therefore 
the definition of critical temperature is arbitrary.
In this work we identify $T_c$ with the temperature at which $|dM/dT|$ is maximum.
The results summarized in Fig.~\ref{Fig:5}  are in agreement with our previous discussion about the
interplay of the fields and $\mu_5$ on chiral symmetry breaking. In particular, from the data in the figure
we notice that increasing the magnitude of the electric field keeping fixed $\mu_5$ results in a lowering of
$T_c$. Then increasing $\mu_5$ results in an increase of $T_c$ towards its zero field value.

We also notice that for $E=B=0$ the line $T_c(\mu_5)$ is very flat,
while increasing the fields the curvature of the critical line increases: this implies
a different relative change of $T_c$ with $\mu_5$ for different values of the fields.
As a matter of fact for the largest values of the fields considered in Fig.~\ref{Fig:5},
namely $eE=eB=10m_\pi^2$, increasing $\mu_5$ from zero to $300$ MeV results in
a relative increase of $\simeq 50\%$, to be compared with $\simeq 8\%$ in the
case $eE=eB=5m_\pi^2$ and with $\simeq 4\%$ in the case $E=B=0$.
The results  of Fig.~\ref{Fig:5} are in qualitative agreement with \cite{Ruggieri:2016lrn},
even if quantitatively the effect of $\mu_5$ on $T_c$ we find in this work is larger than the one
of \cite{Ruggieri:2016lrn} because in that reference the $\mu_5$-dependence has been neglected
in the field-dependent part of the gap equation.

\section{Conclusions}
In this article we have studied, within  a Nambu-Jona-Lasinio model,
quark matter with a background of parallel electric, $E$ and magnetic, $B$, fields,
in presence of a chemical potential, $\mu_5$, conjugated to chiral density.
The main goal of our study has been the computation of the critical temperature for 
(approximate) chiral symmetry restoration, $T_c$, as a function of the external field strengths
and $\mu_5$; our results are summarized in Fig.~\ref{Fig:5} where we plot $T_c$ versus
$\mu_5$ for several values of $E=B$. Inverse magnetic catalysis (IMC) has been taken into account
in this study by  introducing a dependence of the NJL coupling constant on $B$ as
suggested in \cite{Ferreira:2014kpa}.

The role of the electric field is to inhibit chiral symmetry breaking, leading to a lowering of the
critical temperature, while $\mu_5$ is a catalyzer of chiral symmetry breaking
triggering an increase of $T_c$. Therefore we have studied the combined effects of the fields on the one hand,
and the chiral chemical potential on the other hand, on the critical temperature.
A similar problem was considered in \cite{Ruggieri:2016lrn} in which the equilibration of $n_5$ was considered;
the value of $\mu_5$ was then computed self-consistently by the number equation once the value of
$n_5$ at equilibrium was known.
With respect to \cite{Ruggieri:2016lrn}, the novelty of the present study has been the solution of the gap equation
with finite fields and $\mu_5$ beyond the perturbative analysis. Besides we have introduced IMC
in the present study, which instead was not considered in  \cite{Ruggieri:2016lrn}.
We have not solved simultaneously the gap and the number equations, limiting ourselves to treat $\mu_5$
as an external parameter rather than as the result of the equilibration of $n_5$.
The picture obtained in the present study, however, should be robust regardless the fact that $\mu_5$
is considered a free parameter rather than arising from an equilibrium value of $n_5$.

We have computed the evolution of $T_c$ with $\mu_5$ and fields.
The background fields lead to a lowering of $T_c$
because of the combined effect of the IMC induced by the magnetic field,
and the IMC triggered by $E\cdot B$;
we have found that a small value of $\mu_5$ does not change drastically this result, and that the shift of $T_c$
induced by $\mu_5$ is in quantitative agreement with \cite{Ruggieri:2016lrn}.
On the other hand, for larger values of $\mu_5$ the effect on the critical temperature is more important:
it depends on the strength of the external fields, becoming larger with larger fields, as summarized
in Fig.~\ref{Fig:5}. This effect could not be captured by the perturbative solution of \cite{Ruggieri:2016lrn}
because in the latter case the coupling between $\mu_5$ and the fields in the gap equation was lacking.
Looking at data in Fig.~\ref{Fig:5} we notice that  a value of $\mu_5\simeq 300$ MeV
is enough to bring $T_c$ up to the value at zero field, the exact value of $\mu_5$ being dependent on the magnitude
of the external fields.

An improvement of the present work is the study of the equilibration of $n_5$ and
the self-consistent solution of the gap and number equations
going beyond the perturbative analysis of \cite{Ruggieri:2016lrn},
using the formalism intriduced in the present work. Moreover, it would be interesting
to take into account neutral pion condensation induced by anomaly in the picture, as 
it has been studied in \cite{Cao:2015cka}.
We plan to report on this subjects in
the near future.

{\em Acknowledgments}.
The authors would like to thank the
CAS President's International Fellowship Initiative (Grant No. 2015PM008),
and the NSFC projects (11135011 and 11575190). M. R. acknowledges correspondence
with M. Chernodub and X. G. Huang.


\end{document}